\documentclass[article]{aa}
\usepackage{txfonts}
\usepackage{natbib}
\usepackage{graphicx}
\usepackage{color}
\usepackage{longtable,lscape}
\usepackage{float}
\usepackage{subfigure}%

\newcommand{\kmpers}{$\mathrm{km \, s^{-1}}$} 

\newcommand{\cmtwo}{cm$^{-2}$} 
 
\newcommand{\cmthree}{cm$^{-3}$}

\newcommand{\vlsr}{$\upsilon_{\rm LSR}$}              
\newcommand{\vsys}{$\upsilon_{\rm sys}$}

\newcommand{\about}{$\sim$}                       
\newcommand{\expo}[1]{$10^{#1}$}
\newcommand{\texpo}[1]{$\,\times\,10^{#1}$}

\newcommand{\htvao}{H$_2$O}  
\newcommand{\htva}{H$_2$}

\newcommand{\cotvaett}{CO\,(2$-$1)}
\newcommand{\cotretva}{CO\,(3$-$2)}

\newcommand{\amin}{$^{\prime}$}                   
\newcommand{\asec}{$^{\prime \prime}$}
\newcommand{\adeg}{$^{\circ}$}

\newcommand{\atwozero}{$\alpha_{2000}$}
\newcommand{\dtwozero}{$\delta_{2000}$}

\newcommand{\radot}[4]{\mbox{#1$^{\rm h}$#2$^{\rm m}$#3$\stackrel{^{\rm
s}}{_{\bf\cdot}}$#4}}
\newcommand{\decdot}[3]{\mbox{#1$^{\circ}$#2$^{\prime}$#3$^{\prime \prime}$}}
\newcommand{\lsun}{$L_{\odot}$}                          
\newcommand{\msun}{$M_{\odot}$}

\newcommand{\herschel}{\textit{Herschel}}

\newcommand{\apex}{APEX}

\newcommand{\irasfemtontre}{\mbox{IRAS\,15398--3359}}

\titlerunning{A young bipolar outflow from IRAS\,15398-3359}

\begin{document}
   \title{\textbf{\Large{A young bipolar outflow from IRAS\,15398-3359
   }}}


   \author{P. Bjerkeli
          \inst{1,2},
          J.K. J\o rgensen\inst{1}
          \and
          C. Brinch\inst{3}
          }
   \institute{
             Centre for Star and Planet Formation, Niels Bohr Institute \& Natural History Museum of Denmark, University of Copenhagen, {\O}ster Voldgade 5-7, DK-1350 Copenhagen K., Denmark \\
             \email{per.bjerkeli@nbi.dk}
             \and
             Department of Earth and Space Sciences, Chalmers University of Technology, Onsala Space Observatory, 439 92 Onsala, Sweden 
             \and
             Niels Bohr International Academy, The Niels Bohr Institute, Blegdamsvej 17, DK-2100, Copenhagen ¯, Denmark
                          }

   \date{Received 07 September 2015 / Accepted 18 January 2016}

\abstract
   {Changing physical conditions in the vicinity of protostars allow for a rich and interesting chemistry to occur. Heating and cooling of the gas allows molecules to be released from and frozen out on dust grains. These changes in physics, traced by chemistry, as well as the kinematical information allows us to distinguish between different scenarios describing the infall of matter and the launching of molecular outflows and jets.}
   {We aim at determining the spatial distribution of different species, of different chemical origin. This is to examine the physical processes in play in the observed region. From the kinematical information of the emission lines we aim at determining the nature of the infalling and outflowing gas in the system. We also aim at determining the physical properties of the outflow.}
   {Maps from the Sub-Millimeter Array (SMA) reveal the spatial distribution of the gaseous emission toward \irasfemtontre. The line radiative transfer code LIME is used to construct a full 3D model of the system taking all relevant components and scales into account.}
   {CO, HCO$^+$ and N$_2$H$^+$ are detected and is shown to trace the motions of the outflow. For CO, also the circumstellar envelope and the surrounding cloud have a profound impact on the observed line profiles. N$_2$H$^+$ is detected in the outflow, but is suppressed towards the central region, perhaps due to the competing reaction between CO and H$_3^+$ in the densest regions as well as destruction of N$_2$H$^+$ by CO. N$_2$D$^+$ is detected in a ridge south-west from the protostellar condensation and is not associated with the outflow. The morphology and kinematics of the CO emission suggests that the source is younger than \about1000 years. The mass, momentum, momentum rate, mechanical luminosity, kinetic energy and mass-loss rate are also all estimated to be low. A full 3D radiative transfer model of the system can explain all the kinematical and morphological features in the system.}
   {}


\maketitle
%
\section{Introduction}
When stars form, several distinctly different physical components are present in the region, i.e., a protoplanetary disk, a collapsing protostellar envelope, and a (bipolar) molecular outflow. The chemistry in these regions is complex and the abundance of various species varies with the changing physical conditions in time and space. Heating from the protostar and from shocks can evaporate molecules into the gas-phase but molecules can also freeze out in the regions where temperatures are low. 

The Class 0 \citep{Andre:1990fk,Andre:1993fk} protostar \irasfemtontre\ \citep{Shirley:2000qf} is located in the Lupus I cloud \citep[\atwozero~=~\radot{+15}{43}{02}{2}, \dtwozero~=~\decdot{-34}{09}{06.7},][]{Jorgensen:2013lr} at a distance of 155 pc \citep{Lombardi:2008lr}. The source has a bolometric temperature of 44~K \citep{Jorgensen:2013lr} and is known to harbour a molecular outflow \citep[e.g.][]{Tachihara:1996vn,van-Kempen:2009rt}. The region has, however, not attracted much interest until very recently, not least through the ALMA, Cycle 0 observations that were carried out towards this region \citep{Jorgensen:2013lr,Oya:2014kx}. These observations shows that the source most likely underwent a burst in accretion during the last 100 -- 1000 years, which was manifested by an absence of HCO$^{+}$ in the vicinity of the protostellar object. \irasfemtontre\ is also known to show interesting chemical signatures. \citet{Sakai:2011fr} reported an increase in carbon-chain molecules in the inner regions closest to the protostellar source (500 - 1000 AU), and this YSO is one of the so-called Warm Carbon-Chain Chemistry sources. Being one of the very nearby, young outflow sources with an interesting chemistry makes it an excellent target for detailed studies of the gas morphology and kinematics of different species.

The mass-loss, traced by the outflow likely has its origin close to the protostellar object itself \citep[see e.g.][]{Banerjee:2006lr,Machida:2008rt} and it is in fact one of the most spectacular features of the star formation process. The gas is likely ejected through magneto-centrifugal processes \citep[see e.g.][]{Shang:2007yq,Pudritz:2007rp}, however, the details of these mechanisms are not yet fully understood. Accretion shocks are also believed to be present in the inner region, but at present it is difficult to unambiguously determine if infall occurs in any specific object. Self-absorbed, optically thick spectral lines, where the red-shifted component is enhanced with respect to the blue-shifted one, can be a sign of this. However, in order to distinguish between the different processes, one needs to calculate the radiative transfer, taking all different components into account simultaneously. This clearly calls for full 3D radiative transfer modelling, which has to this date not been done. In order to compare such models with observations, however, studies of YSOs at high spatial and spectral resolutions are necessary.

   The aim of this paper is to understand the observed kinematical and morphological information in relation to star formation theory. The outline of the paper is as follow. The observations are described in detail in Sec. 2, and the kinematical and morphological distribution of the detected species are discussed in Sec. 3. In Sec. 4 we analyse the observed line profiles and present a radiative transfer model of the system, taking all different components into account.  The physical properties of the source are discussed in Sec. 5. The main conclusions of the paper are outlined in \mbox{Sec. 6.}
\label{section:introduction}

\section{Observations}
\label{section:observations}
The data for \irasfemtontre\ were obtained from the archive of the Submillimeter Array \citep[SMA;][]{Ho:2004fk}. Parts of these data have previously been utilised in papers by \citet{Chen:2013zr} and \citet{Jorgensen:2015yu}. The data originate from observations on two occasions, in 2009 April 29 and 2009 May 12, covering spectral setups at around 230 and 267~GHz, respectively. During both observations the array was in its compact configuration covering baselines from about 5--87~k$\lambda$ (230~GHz) and 8--109~k$\lambda$ (267~GHz) resulting in beam sizes of 2.4~--~4.5\asec. The observations are sensitive to emission originating on scales smaller than \about20\asec\ \citep{Wilner:1994ph}. The phase centre of the observations was in both cases, \atwozero~=~\radot{+15}{43}{02}{16}, \dtwozero~=~\decdot{--34}{09}{09.0}. The SMA correlator was in both instances set to provide a uniform coverage of 0.812~MHz (0.9--1.1~km~s$^{-1}$) in the twentyfour 104~MHz ``chunks'' distributed over each of its upper and lower sidebands, except in some of the chunks covering specific lines where a higher spectral resolution (0.203--0.406~MHz; 0.22--0.56~kms$^{-1}$) was chosen (Table~\ref{table:correlator}).  
For the analysis presented in this paper, as well as the analysis of \citet{Jorgensen:2015yu} focusing solely on the C$^{18}$O\,(2--1) emission, the data were downloaded from the archive, calibrated using the standard recipes using the IDL/MIR software\footnote{https://www.cfa.harvard.edu/~cqi/mircook.html} and imaged using Miriad \citep{Sault:1995jk}.

\begin{table*}[t]
\flushleft
\caption{Correlator setups}
\resizebox{\hsize}{!}{
\begin{tabular}{llllllll}\hline\hline
  \noalign{\smallskip}
Chunk & Number of channels / Resolution & LSB frequency & Line & $E_{\rm{up}}$\,(K) & USB frequency & Line & $E_{\rm{up}}$\,(K) \\ 
  \noalign{\smallskip}
\hline
  \noalign{\smallskip}
\multicolumn{3}{l}{\textit{230~GHz dataset (219.448 -- 221.430 \& 229.447 -- 231.429~GHz): }} \\
  \noalign{\smallskip}
13 & 512 / 0.203~MHz & 220.343--220.447~GHz & $^{13}$CO\,(2 -- 1) & 15.9 &230.430--230.534~GHz & $^{12}$CO\,(2 -- 1)& 16.6 \\
14 & 256 / 0.406~MHz & 220.261--220.364~GHz & - & - &230.513--230.616~GHz & $^{12}$CO\,(2 -- 1) & 16.6\\
23 & 512 / 0.203~MHz & 219.529--219.633~GHz &  C$^{18}$O\,(2 -- 1) &15.8 &231.244--231.338~GHz & N$_2$D$^+$\,(3 -- 2) & 22.2\\
 \noalign{\smallskip}
\multicolumn{3}{l}{\textit{267~GHz dataset (267.499 -- 269.522 \& 277.559 -- 279.562~GHz):} } \\
 \noalign{\smallskip}	
24 & 512 / 0.203~MHz & 267.499--267.603~GHz & HCO$^+$\,(3 -- 2) & 25.7 &279.458--279.562~GHz & N$_2$H$^+$\,(3 -- 2) & 26.8 \\ \hline
\end{tabular}
\label{table:correlator}
}
\end{table*}

In addition to the Submillimeter Array data we also utilise ALMA Cycle~0 observations of C$_2$H from Early Science program 2011.0.00628.S, also previously presented by \citet{Jorgensen:2013lr}. Those observations provide maps of the C$_2$H $N=4-3$ emission at 349.4~GHz with an angular resolution of approximately 0.5\asec. For further details about the ALMA data we refer to that paper.

\section{Results}
\label{section:results}   
\subsection{Kinematics}
\begin{figure}[]
   \rotatebox{0}{\includegraphics[width=0.53\textwidth]{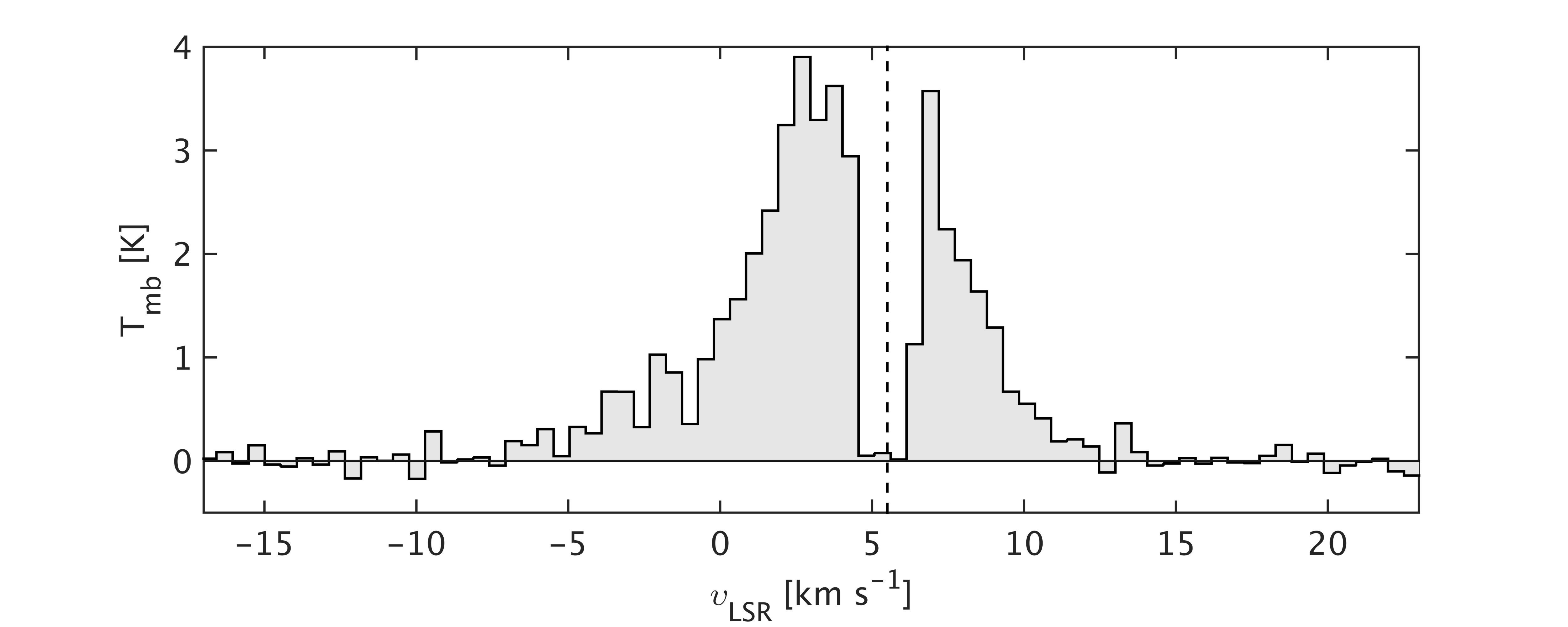}}
      \caption{$^{12}$CO\,(2--1) spectrum towards the central position. The velocity of the source, \vlsr~=~+5.5~\kmpers, is indicated with a dashed vertical line.}
               \label{fig:center12co}
   \end{figure}

\begin{figure*}[]
   \centering
   \vspace{-1cm}
   \hspace{-1.10cm}
    \rotatebox{270}{\includegraphics[width=0.75\textwidth]{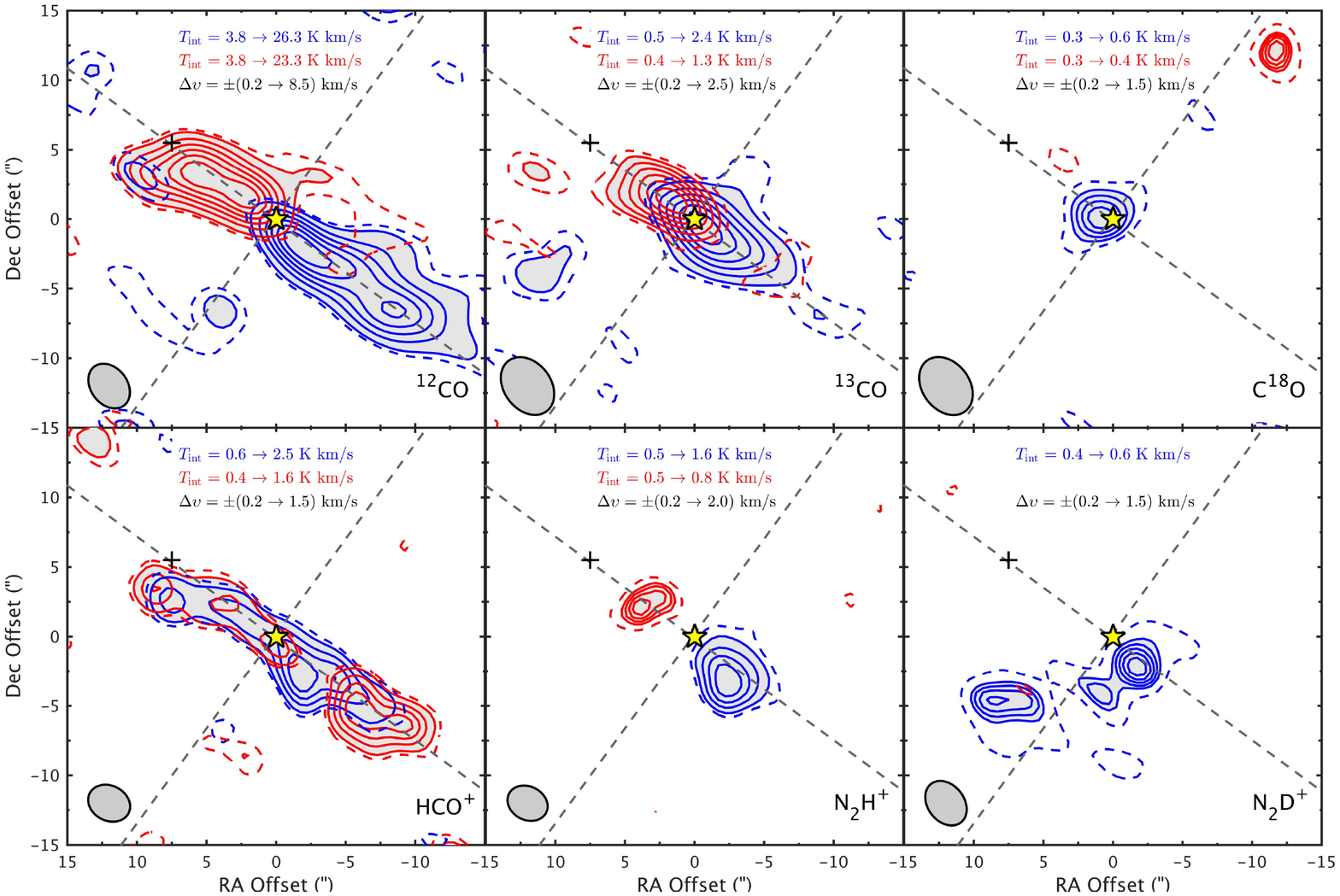}}
      \vspace{-0.7cm}
      \caption{Integrated emission in the red and blue outflow lobes. The dashed lines mark the 2 $\sigma$ levels and the first thick line mark the 3 $\sigma$ levels. In each panel, the value for the first and last contour in the shaded region are given, together with the velocity intervals over where the emission has been integrated. The beam size and molecules are marked in the lower corners of each panel. Dashed grey lines indicate the cut for the position-velocity diagram presented in Fig.~\ref{fig:pv1}. The location of the central source is indicated with a yellow star and the location of the outflow position discussed in Sec.~\ref{sec:contributiontoemissionlineprofilesfromthedifferentcomponents} is indicated with a plus sign. The maps are centered at \atwozero~=~\radot{+15}{43}{02}{2}, \dtwozero~=~\decdot{--34}{09}{06.7}.}
         \label{fig:speciesmaps}
   \end{figure*}

\begin{figure*}[]
   \centering
   \includegraphics[width=1.15\textwidth]{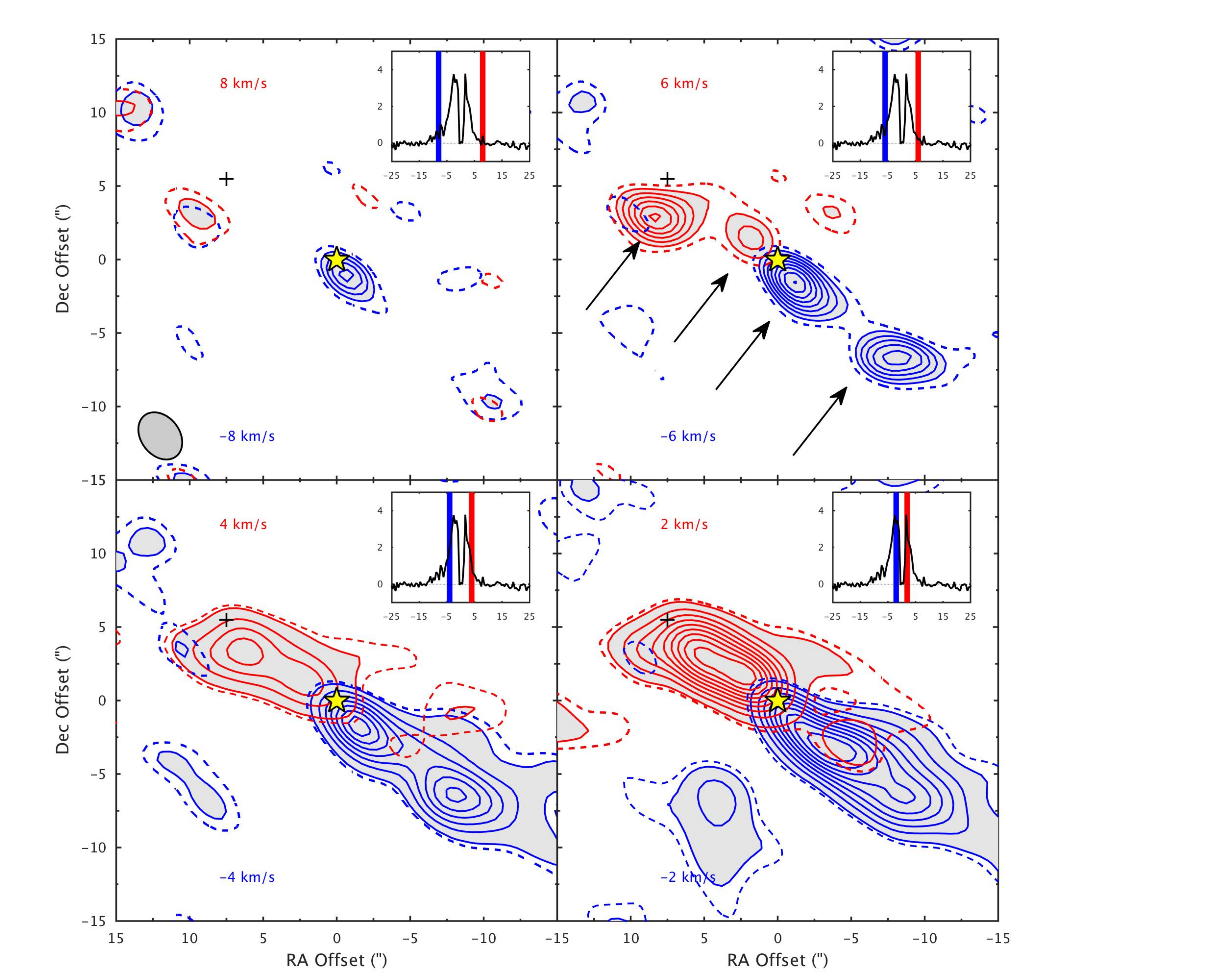}
      \caption{Integrated $^{12}$CO emission in different velocity intervals, where $\Delta \upsilon$~=~2~\kmpers (inset in each panel). The centre velocities ($\upsilon - \upsilon_{\rm{sys}}$) for each bins are indicated in the respective panel. High-velocity gas ($\Delta \upsilon$~$>$~5~\kmpers) is detected in four different spots along the outflow axis (indicated by black arrows in the upper right panel). First solid thick contour is at 3$\sigma$ and the 2$\sigma$ level is indicated with a dashed line.}
         \label{fig:cochannelmap}
   \end{figure*}

\begin{figure}[]
   \centering
   \includegraphics[width=0.53\textwidth]{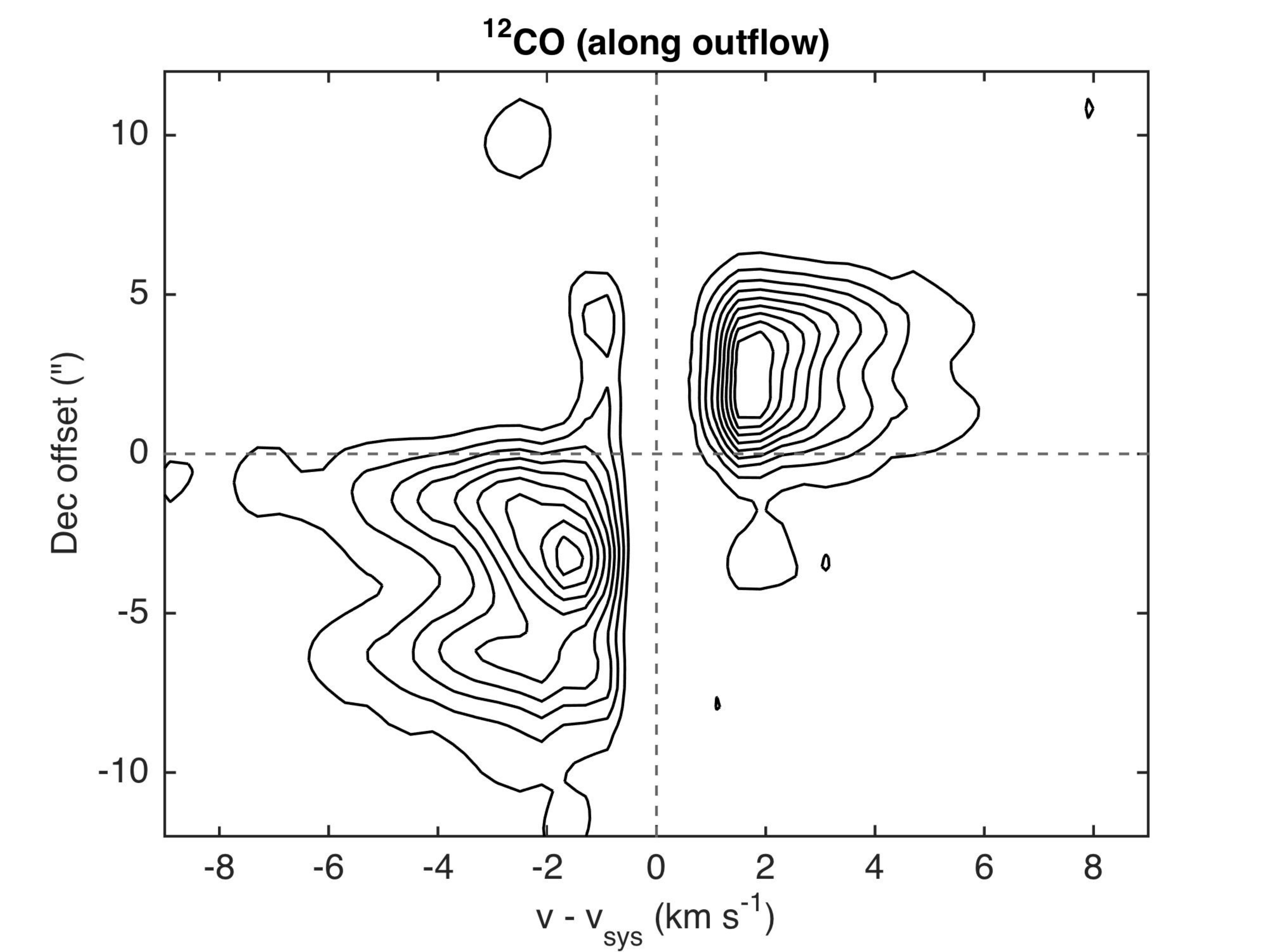}
   \includegraphics[width=0.53\textwidth]{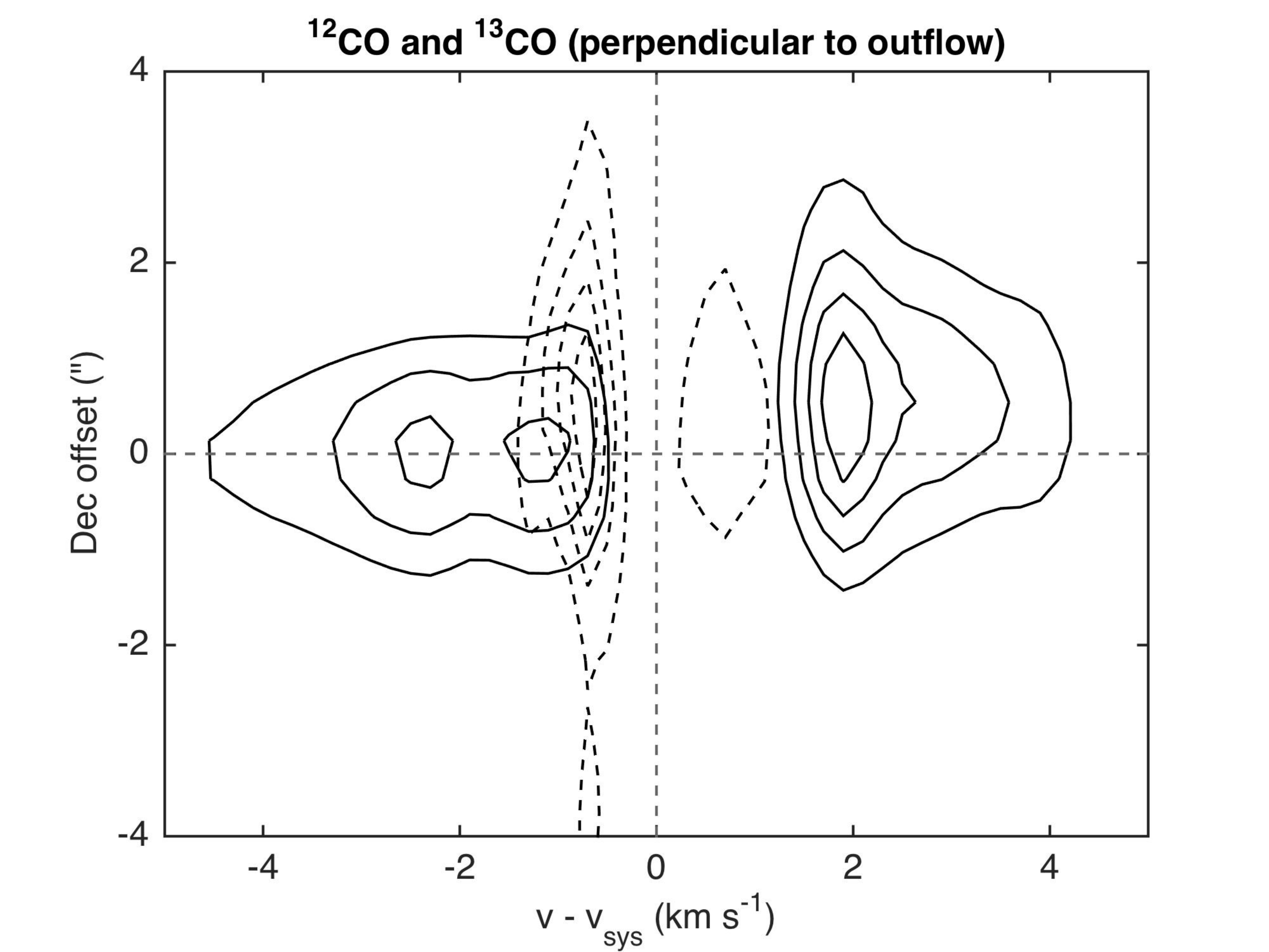}
      \caption{Position velocity diagram along the outflow direction (upper panel) and along the direction of the presumed disk structure (lower panel). Solid contours are for $^{12}$CO and dashed contours are for $^{13}$CO. The cut along the outflow direction reveals episodic ejection events and weak signs of acceleration along the outflow axis. $\upsilon - \upsilon_{\rm{sys}}$ and the position of the source are indicated with dashed grey lines. Note that the ordinate shows the offset in Declination from the central source.}
         \label{fig:pv1}
   \end{figure}

The $^{12}$CO emission emanating from \irasfemtontre\ was detected at a high signal-to-noise ratio. The observed $^{12}$CO line profiles show a triangular shape with a central absorption at the systemic velocity, $\upsilon_{\rm{sys}}$~=~+5.5~\kmpers\ (determined by fitting Gaussians to the observed spectra), indicative of infall and/or outflow emission. This spectrum is presented in Fig.~\ref{fig:center12co}, where the velocity is with respect to \vlsr. From here on and throughout this paper, however, all velocities are reported with respect to $\upsilon - \upsilon_{\rm{sys}}$ for clarity. 
In the other CO isotopologues, and also in several of the other observed species, strong outflow activity  and clear blue- and red-shifted asymmetries are observed. The velocity integrated spatial distributions of the detected lines are shown in Fig.~\ref{fig:speciesmaps}. 
Compared to the 1--0 transition of CO, the 2--1 transition is particularly well suited for kinematical studies (see Sec.~\ref{section:discussion}) due to its higher upper state energy ($E_{\rm{up}}$~=~16.6~K), which makes it less sensitive to the low-temperature quiescent gas. High-velocity emission (up to \about8~\kmpers\ offset from the systemic velocity) is detected both in the outflow and towards the central source. In Fig.~\ref{fig:cochannelmap}, 
we present a channel map of the region in $^{12}$\cotvaett. This figure shows the velocity integrated emission in velocity intervals of 2~\kmpers. The two upper panels show the gas moving at higher velocities compared to the systemic velocity ($\Delta \upsilon$~$>$~5~\kmpers) while the gas at lower velocities ($\Delta \upsilon$~$<$~5~\kmpers) is presented in the lower panels. In Fig.~\ref{fig:pv1}, a position velocity cut along the direction of the outflow (PA = 35\adeg) and the presumed disk-like structure \citep{Oya:2014kx} is presented. 
Other species are only detected at a low velocity within $\Delta\upsilon\,\simeq$~3~\kmpers\ from systemic velocity. The mass loss is clearly detected also in the $^{13}$CO maps, whilst the C$^{18}$O emission is only detected at blue-shifted velocities, closest to the central source. HCO$^+$ is detected in the outflow as well, at relatively low radial velocities ($\Delta \upsilon$ \about3~\kmpers) compared to \vsys\ and both red-shifted and blue-shifted emission is detected in both outflow lobes. N$_2$H$^+$ is detected in two knots on each side of the protostar and the lines are narrow, i.e., a few \kmpers. The knots are, however, clearly separated in velocity.
 \subsection{Morphology}
\label{section:distributionofdetectedspecies}
The $^{12}$CO emission is confined to the two outflow lobes emanating from \irasfemtontre, and the morphology of the flow is consistent with recent ALMA observations presented in \citet{Jorgensen:2013lr} and \citet{Oya:2014kx}. In those observations, the outflow cavities are clearly detected in the emission from C$_2$H \citep[][their Fig.~1]{Jorgensen:2013lr}, suggesting the presence of a wide-angle wind \citep[see e.g.][]{Lee:2000zr}. The spatial resolution of the \cotvaett\ observations presented here (\about3\asec) is, however, too poor to reveal such variations in the emission on small spatial scales. Consequently, the absence of HCO$^+$ emission towards the central source \citep{Jorgensen:2013lr} is not evident from this dataset either, since the beam size is a factor of \about10 larger than in the ALMA observations. Also, the optical thickness is expected to be higher in the data presented here. The HCO$^+$ peak positions in the outflows are fairly well correlated (within a few arc seconds) with the positions of the $^{12}$CO emission maxima at high velocity (see Sec. 4.1). The spatial resolution of the dataset does not allow us to tell whether the small spatial differences between CO and HCO$^+$ are due to chemistry or not. One could envision a scenario where HCO$^+$ is partially destroyed in shocked spots along the outflow \citep{Podio:2014gb}. The presence of red-shifted and blue-shifted emission in both outflow lobes could suggest that the HCO$^+$ emission stems from the cavity walls \citep[c.f.][]{Tappe:2012lh} where the velocities perpendicular to the outflow axis are expected to be largest. This origin, combined with the fact that the outflow only has a small inclination with respect to the plane of the sky \citep[$i$~=~20\adeg, see Sec.~\ref{section:averyyoungoutflowsource} \&][]{Oya:2014kx}, can explain the observed emission.

The N$_2$H$^+$ emission traces the outflow with red-shifted emission in the northeastern lobe and blue-shifted in the southwestern lobe. No emission is detected towards the protostar and the peak positions are located where the $^{12}$CO emission peaks at higher velocities (see Fig.~\ref{fig:cochannelmap}). At present, we can not with certainty determine the origin of the N$_2$H$^+$ emission. N$_2$H$^+$ is not an obvious shock tracer and it has only very recently been observed in shocked regions \citep[i.e. L1157-B1,][]{Codella:2013uq,Podio:2014gb}. The scenario, where N$_2$H$^+$ is not detected towards the protostar is morphologically reminiscent to the L\,483 case, where the abundance of N$_2$H$^+$ in the dense central region is reduced due to reactions with gas phase CO \citep{Jorgensen:2004lr,Jorgensen:2004rt}. Also in the maps presented in this paper, the C$^{18}$O emission peaks close to the protostar whilst the N$_2$H$^+$ emission peaks in the outflow component where the CO column density is expected to be lower.

N$_2$D$^+$ is detected in ridge to the South-southeast from \irasfemtontre\ and only at blue-shifted velocities. The spatial distribution is not well correlated with the extent of the outflow, neither the protostellar envelope. We can therefore not exclude the possibility that this gas instead is associated with the gas surrounding \irasfemtontre.

\section{Analysis}
\subsection{Observed line profiles}
\label{section:observedlineprofileshapes}
The observed line shapes give important clues, when it comes to the relative contribution to the emission from different components. 
The outflow morphology is clearly evident in the emission from \irasfemtontre. However, the emission at velocities within a few \kmpers\ from $\upsilon_{\rm{sys}}$ could also, in addition to the outflow component, have a contribution from the protostellar envelope, and/or the protoplanetary disk. Not to forget, the surrounding cloud can have significant impact at the lowest velocities. Inspection of the line profiles in $^{12}$CO shows a prominent absorption feature at the systemic velocity. The reason for this absorption feature could be several. An infalling envelope can give rise to a line profile where the blue-shifted wing component is enhanced with respect to the red-shifted one. A search for infall signatures in the emission from H$_2$CO and CS was presented in \citet{Mardones:1997fk}. This study does not reveal any detectable inflow of gas, however, in a more recent study by \citet{Kristensen:2012lr}, the \htvao\ line profiles observed with \herschel-HIFI clearly show infall signatures. The \htvao\ line profiles were further analysed in a study by \citet{Mottram:2013qy}, where the data is consistent with an infall rate of \about3\texpo{-5}~\msun\ yr$^{-1}$. In the \irasfemtontre\ case, the central absorption could also be affected by the interferometer filtering out the emission on large scales ($>$20\asec, see Sec.~\ref{section:observations}), i.e., the low velocity emission. Arguing against this is the smooth distribution of the gas indicating that we recover most of the emission also at low velocities with respect to \vsys\ \citep[cf. the clumpy distribution in the channel maps presented by][]{Arce:2013fj}. Even though the lack of single-dish data prevents us from putting quantitative numbers on the effect of loss of short-spacings, we still find it unlikely that this will significantly affect the emission in the line wings. This emission is believed to originate in the compact regions, either where the jet impact the surrounding medium, and/or in the cavity walls of the flow/bow-shock.

\subsection{Radiative transfer modeling}
\label{section:radiativetransfermodeling}
To investigate to which extent each component contributes to the observed $^{12}$CO\,(2--1) emission lines at various velocities, we construct a 3D model of \irasfemtontre, using the Monte-Carlo code LIME \citep{Brinch:2010rm}. 
\begin{figure}[]
   \flushleft
      \includegraphics[width=0.535\textwidth]{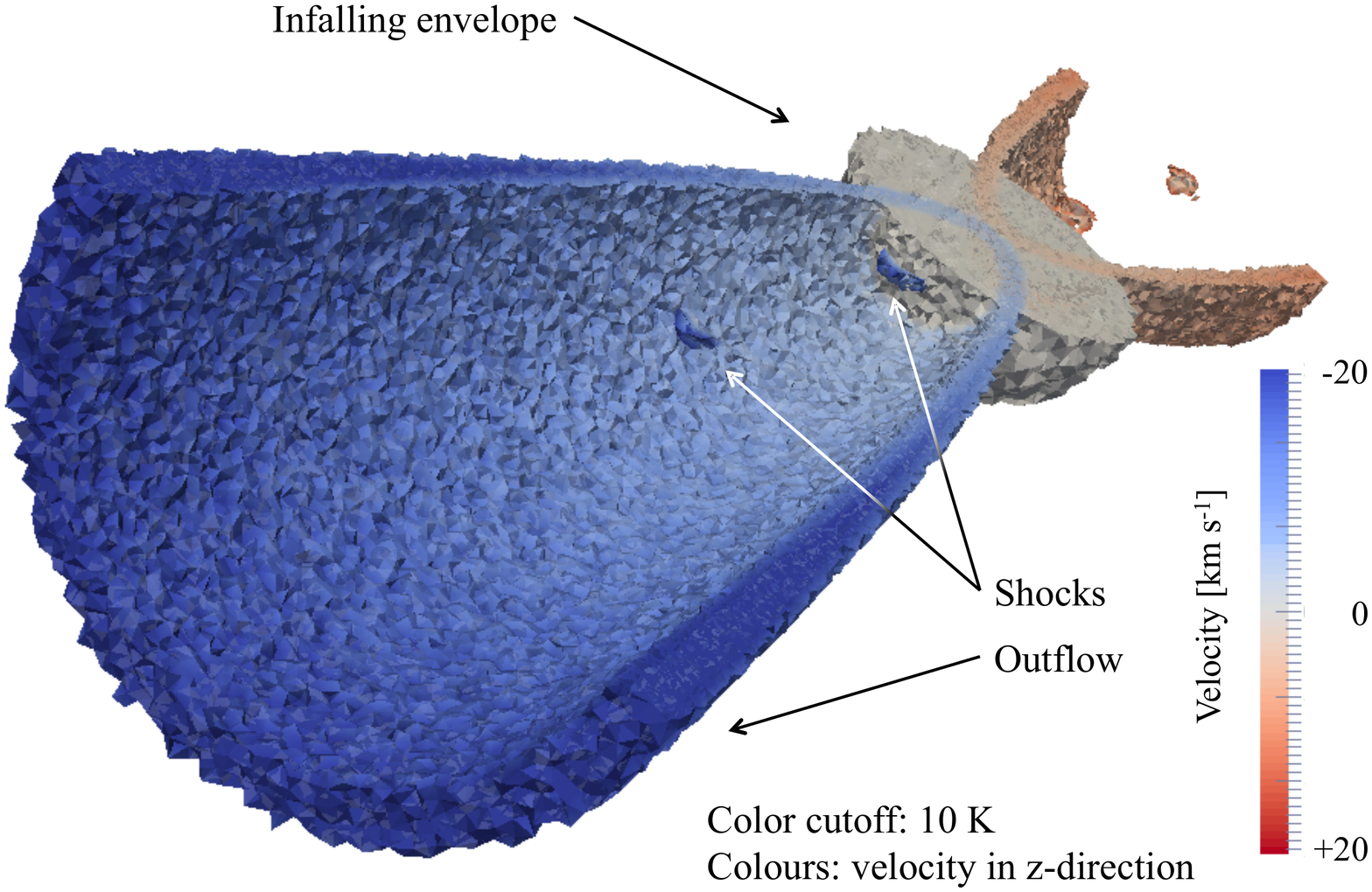}
      \caption{A cut through the radiative transfer model described in the text (partially transparent). The model is viewed from the South-West direction for clarity. Included in the figure, is the molecular outflow, shocked regions along the jet axis and infalling envelope. Colours show the velocity field in the outflow direction in the regions where the temperature is higher than 10~K. The surrounding cloud component, which is also included in the modeling, is not presented in this figure.}
         \label{fig:limemodel}
   \end{figure}
This radiative transfer code does not put any constraints when it comes to the complexity of the models. The density, temperature, abundance, and velocity structures in the different components are given as analytical descriptions to the LIME code. The irregular Delaunay grid used by LIME is generated by random sampling of the input model. However, the sampling probability is weighted by the density and temperature structure of the model, resulting in a finer grid in the regions with the strongest emission. Once the grid is calculated, LIME starts the iterative process to calculate the rotational level populations. The CO data file that is used was downloaded from the LAMDA\footnote{http://home.strw.leidenuniv.nl/$\sim$moldata/} database \citep{Schoier:2005fk}, where the collisional rate coefficients are taken from \citet{Yang:2010vn}. The ortho-to-para ratio for \htva\ is assumed to take the thermalized value, 3. When convergence is reached, the model is ray-traced to produce fits files that can be compared to the observations. The image resolution for these fits files is set to 0.25\asec\ which is equivalent to 40~AU at the distance of \irasfemtontre. The detailed geometry of the model is discussed in the sections below.
\subsubsection{The model components}
\label{sec:themodelcomponents}
Included in the model are the components that have been proposed to be the major origins of the observed emission from protostellar regions, i.e., infalling envelope, surrounding cloud, outflow cavity and spot shocks originating in the jet. Although the possibility of a circumstellar disk exist, we do not include this component in the model since these observations are not sensitive to the emission at these spatial scales. A rotationally supported disk is hinted by the H$_2$CO emission observed with ALMA \citep{Oya:2014kx}. However, the radius of this feature is likely smaller than \about200~AU, i.e., less than the angular resolution obtained here, viz. \about400~AU. It should also be mentioned that no velocity gradient perpendicular to the jet direction is detected in the dataset presented in this paper.

The structure of the infalling envelope is based on the best-fit model presented in \citet{Mottram:2013qy} and has a radius, $R_{\rm{env}}~=~$4900~AU. The velocity field follow a power-law,
\begin{equation}
\upsilon = \upsilon_0 \left( \frac{r}{r_{\upsilon_0}} \right)^{-p_{\upsilon}}, 
\end{equation}
where the exponent $p_{\upsilon}$ is set to 0.5 and $r_{\upsilon_0}$ is set to 1000 AU. Furthermore, the density profile is described by, 
\begin{equation}
n = n_0 \left( \frac{r}{r_{n_0}} \right)^{-p_{n}},
\end{equation}
where the exponent $p_{n}$ is set to 1.4, $r_{n_0}$ is set to 6.1~AU, and $n_0$ is set to 2\texpo{9}~\cmthree. These values are taken from \citet{Mottram:2013qy}. However, in this case, the velocity $\upsilon_0$ at 1000 AU is set to a slightly lower value, viz. 0.1~\kmpers. Although a higher infall velocity provide a better fit to the width of the line, the larger gradient will also decrease the optical depth in the line, leading to significantly stronger emission than what is observed. Similarly as for the velocity and density structures, the temperature at 175~AU is set to 30~K and follows a power-law with an exponent of 1.5 \citep{Jorgensen:2013lr}.

The outflow is described by a wind-driven shell \citep[][their Fig.~21]{Lee:2000zr} and has an inclination angle of, $i$~=~20\adeg\ \citep{Oya:2014kx} with respect to the plane of the sky. It is radially expanding with a Hubble-law velocity structure, and in cylindrical coordinates, the structure and velocity of this shell, is described by,
\begin{equation}
z = CR^2,~\upsilon_R = \upsilon_0\,R,~\upsilon_{z} = \upsilon_0\,z,
\end{equation}
where $z$ is the distance in the direction of the outflow, $R$ is the radial extent perpendicular to the outflow axis, $\upsilon_R$ is the velocity in the radial direction and $\upsilon_z$ is the velocity in the direction of the outflow axis. $\upsilon_0$ and $C$ are free parameters that here are set to 1~\kmpers~arcsec$^{-1}$ and 1~arcsec$^{-1}$ to fit the observed velocity extent and geometry of the outflow. The choice of geometry is supported by the PV diagram of H$_2$CO\,(5$_{15}$- $4_{14}$) where an elliptic structure and expansion is obvious \citep{Oya:2014kx}. 
\begin{figure}[]
   \centering
  \includegraphics[width=0.50\textwidth]{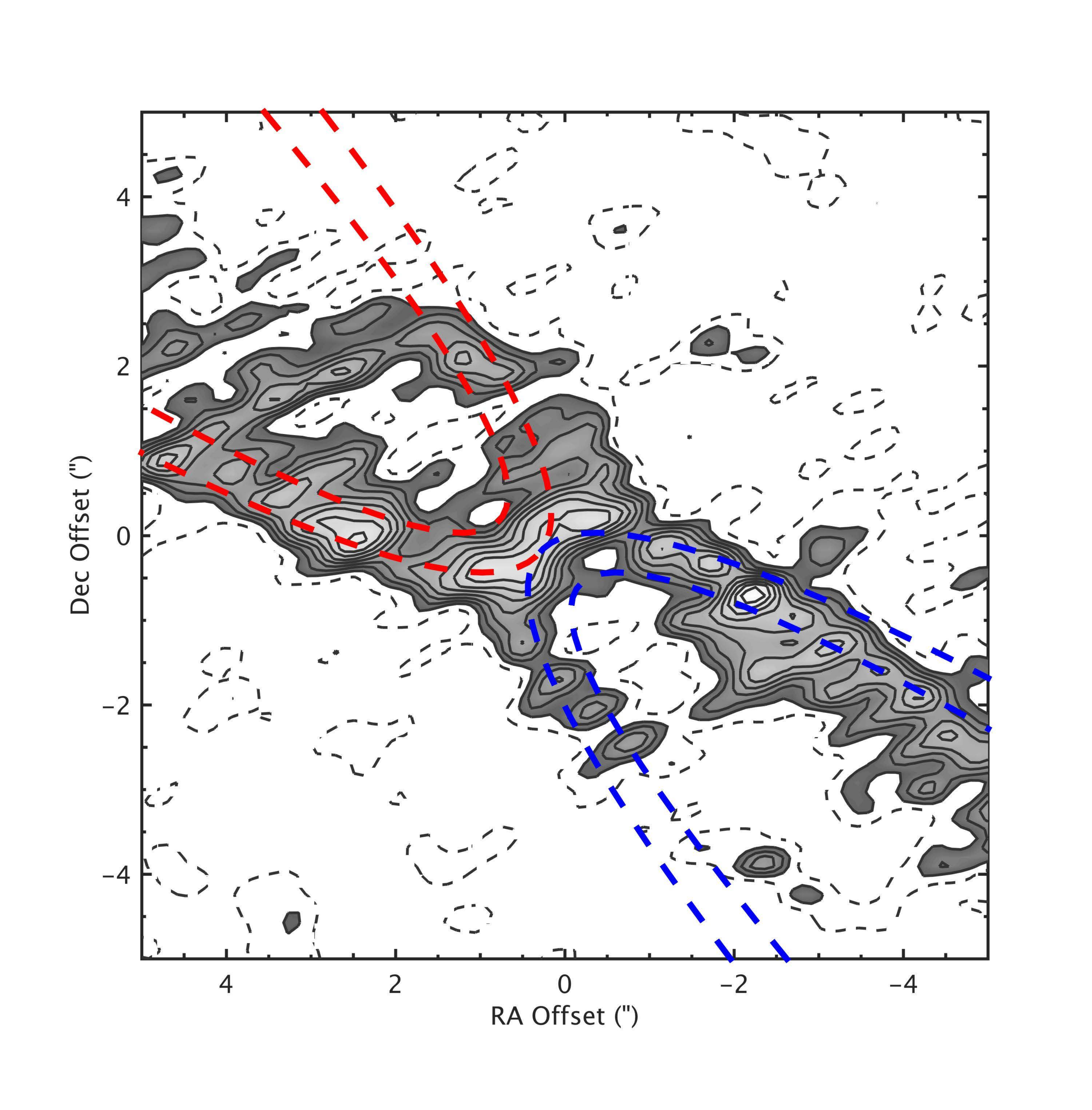}
      \caption{The model of the outflow cavity (red and blue dashed lines) over plotted on the C$_2$H emission acquired with ALMA. The coloured region is where the signal is stronger than 3$\sigma$ and the 1$\sigma$ level is indicated by dashed lines. }
         \label{fig:cchandlime}
   \end{figure}
The thickness of the shell is taken from the observed extent of the C$_2$H emission and is set to 100~AU (see Fig.~\ref{fig:cchandlime}). The velocity increases from the inner to the outer region and is highest at the centre of the shell and drops to half at the edges. The length of the blue- and red-shifted outflow cones ($L_{\rm{blue}}$ and $L_{\rm{red}}$) are 2350 and 1400~AU, respectively. 

In addition to this, bow shocks are added to the inner region of the outflow. The shape of the shocks in cylindrical coordinates follows $z \propto R^{2.4}$ and the velocity decrease with increasing distance from the bow apex \citep{Lee:2000zr}. The distances to the bow-shocks are identical to the distances to the peak positions observed in the CO emission at higher velocities (see upper right panel of Fig.~\ref{fig:cochannelmap}). The separation between each shocked region is 800~AU, with the first two located 600~AU away from the central source.

The surrounding cloud component is represented by a spherical shell surrounding the entire model. This shell has a thickness, $l_{\rm{shell}}~=~$5000~AU, and the temperature is set to 10~K. A visualisation of the model is presented in Fig.~\ref{fig:limemodel}.

\subsubsection{Contribution to the emission line profiles from different components}
\label{sec:contributiontoemissionlineprofilesfromthedifferentcomponents}
Given the large number of free parameters and the considerable amount of CPU time needed, a full chi-square analysis is impracticable. Also, such a study would be of little interest due to the many parameters that can be varied. Models that successfully reproduce the observed line profiles, such as the one presented in this section should always be considered \textit{as one possible solution} to the problem. The aim of this study is, therefore, not to reproduce the exact shape of the line profiles in every position of the map, but instead to understand the contribution from each component when taking the most prominent spectral features into account. 
\begin{figure*}[]
   \centering
   \begin{tabular}{c c }    
 \hspace{-1.7cm} M1 (envelope)  &  \hspace{-2.1cm}M2 (outflow)  \\
    \noalign{\smallskip}
   \hspace{-1.7cm} \rotatebox{270}{\includegraphics[width=0.425\textwidth]{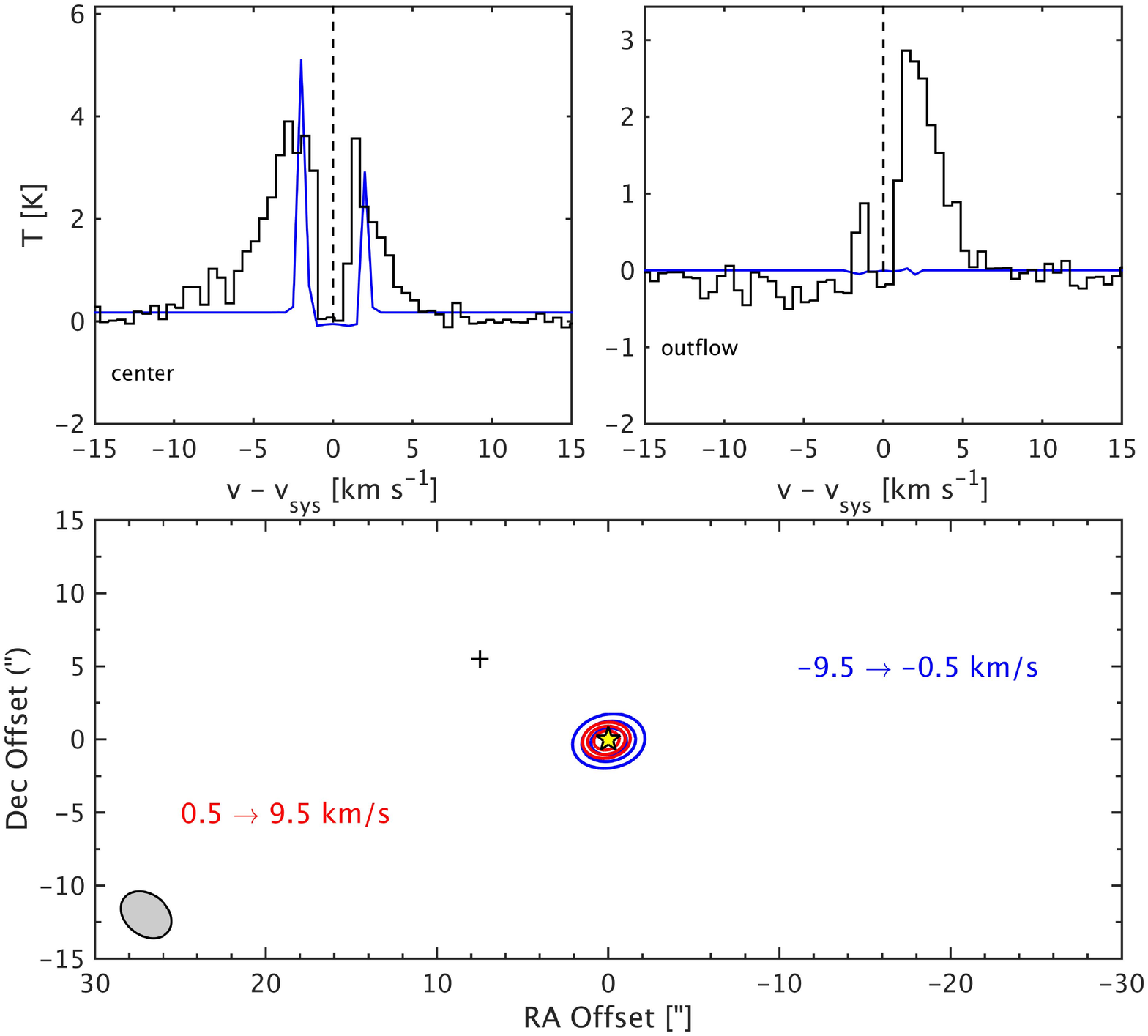}} &  \hspace{-2.1cm} \rotatebox{270}{\includegraphics[width=0.425\textwidth]{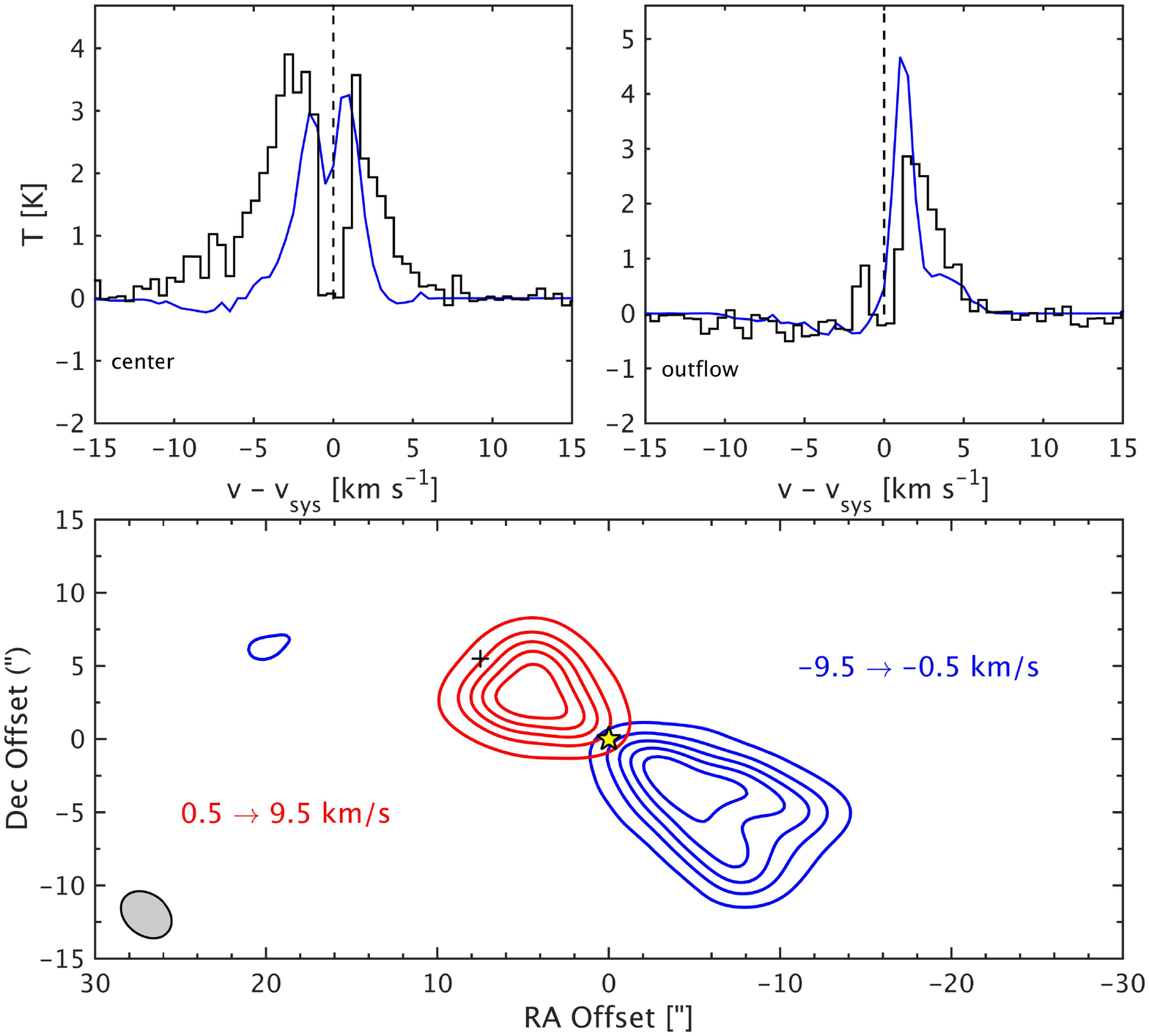}} \\
  \noalign{\smallskip}
 \hspace{-1.7cm} M3 (outflow, envelope \& cloud) &  \hspace{-2.1cm}M4 (outflow, envelope \& shocks) \\
    \noalign{\smallskip}
 \hspace{-1.7cm}  \rotatebox{270}{\includegraphics[width=0.425\textwidth]{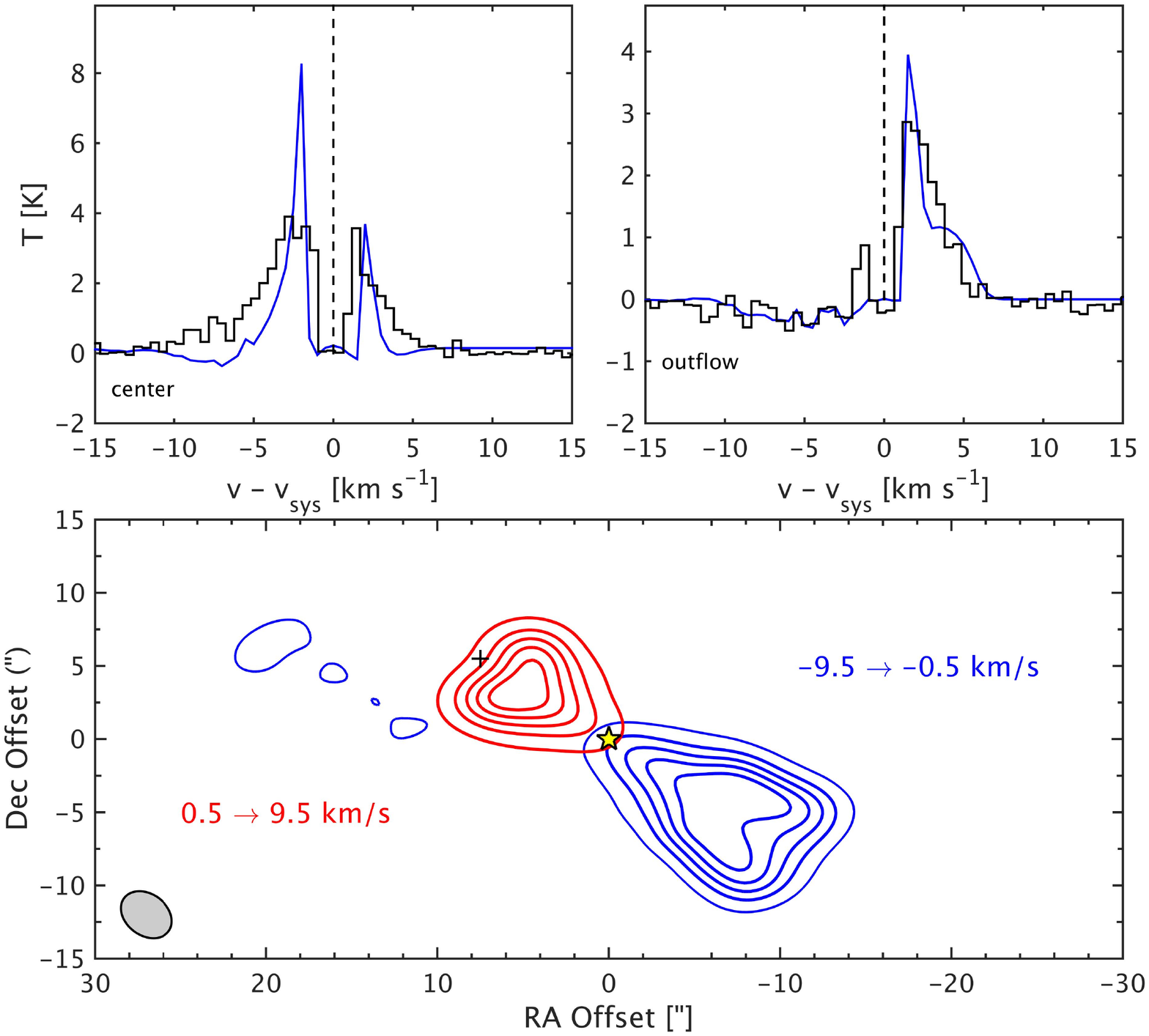}} &  \hspace{-2.1cm} \rotatebox{270}{\includegraphics[width=0.425\textwidth]{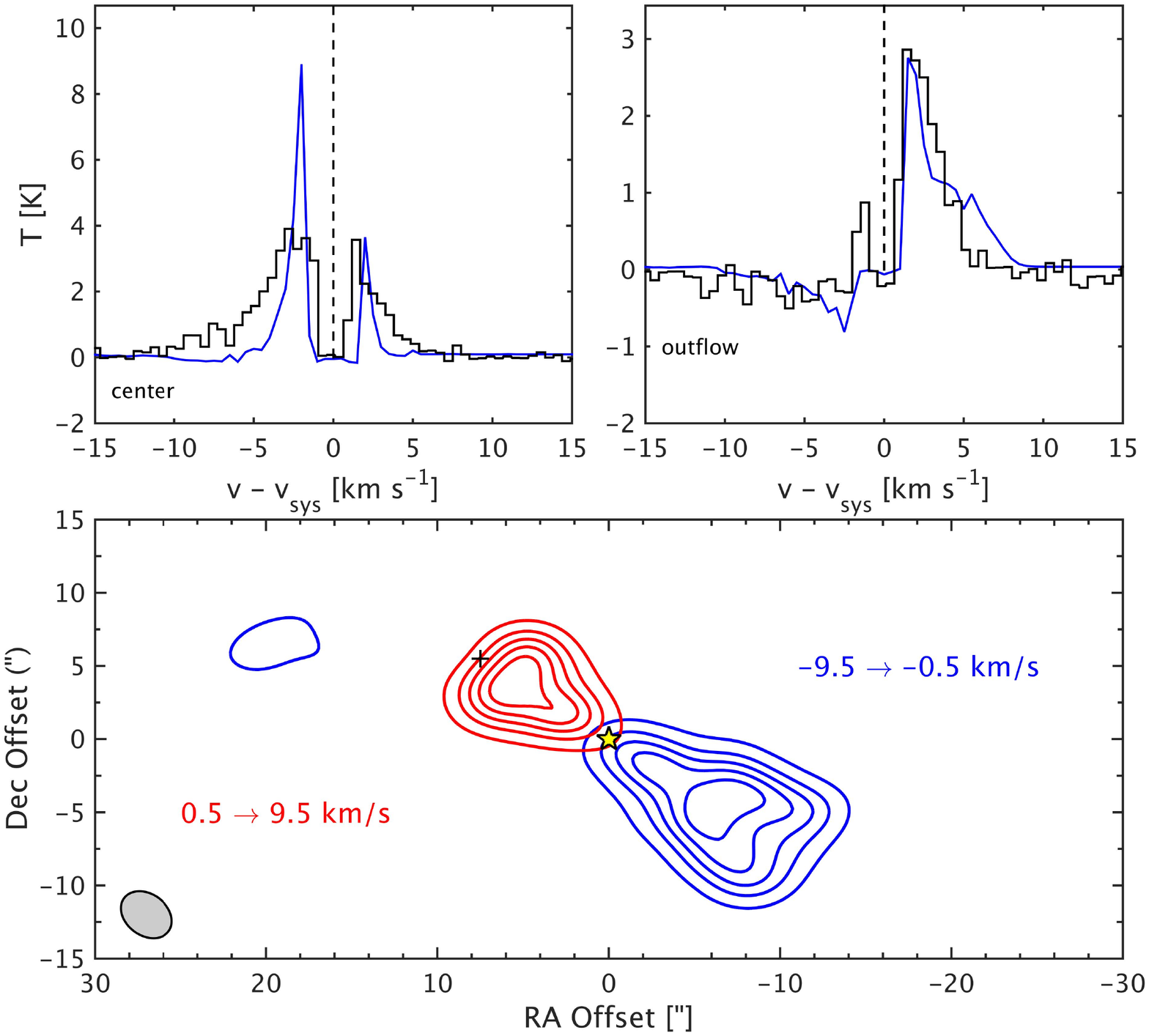}} \\
   \end{tabular}
      \caption{The 4 different LIME models discussed in the text (M1 in the upper left corner and M4 in the lower right). In each panel, the spectrum towards the central region and one outflow position is presented. The observed spectra are plotted as black histograms while the modelled spectra are plotted with blue continuous lines. A contour plot is also presented for each model, where the integrated intensity has been calculated over the red and blue line wings, respectively. The beam size is indicated with a grey ellipse.}
         \label{fig:limemodels}
\end{figure*}   
The method we use here is such that we calculate the radiative transfer for a set of models where the contribution from each component is varied between each run. For the outflow component, the velocity field ($\upsilon$), the temperature ($T$), and density ($n_{H_2}$), are varied in a step by step manner. The same approach is used for the shocked regions and the surrounding cloud. The CO abundance is set to 1\texpo{-4} in all components in the model \citep[see e.g.][]{van-Dishoeck:1987mz}. For each model, 250 000 grid points are used and the number of iterations is set to 16 to ensure convergence of the level populations. The synthetic images of the $^{12}$CO emission, are sampled with the observed visibilities, inverted, cleaned and finally restored, using Miriad. We thus treat the LIME generated fits cubes in the same way as the observed data. This gives us the short-spacing filtered images that can be directly compared to the observations. In Fig.~\ref{fig:limemodels}, we present the four different models (M1 to M4) that fit the observations best, when allowing different components (envelope, outflow, shocks, surrounding cloud) to contribute to the emission. The spectrum towards the central position and one outflow position ($\Delta$RA~=~+7.5\asec, $\Delta$Dec~=~+5.5\asec), as well as a map of the emission (for each individual model) in the blue- and red-shifted velocity ranges as indicated in the figure. The physical input parameters are summarised in Table~\ref{table:limemodels}.
\begin{table}[t]  
\flushleft
\caption{Parameters in the models discussed in the text and presented in Fig.~\ref{fig:limemodels}.}              
\label{table:limemodels}      
\begin{tabular}{l l l l l l l l l l l l l l l l l l l l}          
  \hline\hline                        
  \noalign{\smallskip}
   \noalign{\smallskip}
 \textit{Outflow parameters:} \hspace{2.75cm} \\ 
  \noalign{\smallskip}
Inclination angle ($i$): & 20\adeg \\
Temperature ($T$): & 100 K \\
Structure parameter of the shell ($C$): & 1.0 arcsec$^{-1}$\\
Velocity parameter of the shell ($\upsilon_0$): & 1.1 \kmpers~arcsec$^{-1}$\\
Length of lobe ($L_{\rm{blue}}$): & 2350~AU \\
Length of lobe ($L_{\rm{red}}$): & 1400~AU \\
   \noalign{\smallskip}
    \noalign{\smallskip}
    \textit{Envelope parameters:} \\
     \noalign{\smallskip}
Radius of envelope ($R_{\rm{env}}$): & 4900~AU \\
Velocity at $r_{\upsilon_0}$ ($\upsilon_0$): &0.1~\kmpers \\
Radius where, $\upsilon = \upsilon_0$ ($r_{\upsilon_0}$): & 1000~AU \\
Velocity exponent ($p_{\upsilon}$): & 0.5 \\
\htva\ density at $r_{n_0}$ ($n_0$): & 2\texpo{9}~\cmthree \\
Radius where, $n = n_0$ ($r_{n_0}$): & 6.1~AU \\
Density exponent ($p_n$): & 1.4\\
Temperature at $r_{T_0}$ ($T_0$): & 30~K \\
Radius where, $T = T_0$ ($R_{T_0}$): & 175~AU \\
Temperature exponent ($p_T$):& 1.5 \\
   \noalign{\smallskip}
    \noalign{\smallskip}
    \textit{Shock parameters:} \\
     \noalign{\smallskip}
     Maximum velocity ($\upsilon_{\rm{max}}$): & 20~\kmpers \\
     Temperature ($T$): & 1000~K \\
     \htva\ density ($n_{\rm{H_2}}$): & 1\texpo{8}~\cmthree \\
     Shock distance from source ($d_{\rm{shock}}$): & 600 and 1400~AU \\
     \noalign{\smallskip}
    \noalign{\smallskip}
    \textit{Surrounding cloud parameters:} \\
     \noalign{\smallskip}
     \htva\ column density ($N_{\rm{H_2}}$): & 1\texpo{20}~\cmtwo \\
     Temperature ($T$): & 10~K \\
     Shell thickness ($l_{\rm{shell}}$): & 5000~AU \\
     
   \noalign{\smallskip}
  \noalign{\smallskip}
  \hline                                             
\end{tabular}
\end{table}
In the first model (M1), a pure infall case is considered and no outflow component is present. It is clear that this type of model cannot fully explain the observed emission line profiles. Apart from the obvious fact that no emission is present in the offset position, the line wing strength also decreases with increasing distance from the central source. This is in sharp contrast to the fact that the line wing profiles do not change significantly with increasing distance from the central source. 

On the other hand, a pure outflow scenario (where no infall of gas is present) cannot explain the observed line profiles either. This scenario is presented in M2, and in this case, the emission at the systemic velocity is significant compared to the observations, where the strong absorption is clear. In addition, the shape of the lines towards the central region does not resemble the observed ones. We have not been able to reproduce the observed line profiles by adding a cloud component that is filtered out by the interferometer. This is, however, of little interest since the pure outflow scenario is not very likely. 

A combination of the first two scenarios, can satisfactorily explain the observed emission on both small and large scales (M3). The reason for the strong central reversal of the line profile is in this case the optically thick infalling gas in the central region, absorbing the emission at these frequencies. Included in this model is also a surrounding cloud, where the \htva\ column density is 1\texpo{20}~\cmtwo. It is clear, however, that this component makes only a small contribution to the line profiles, as is also evident when comparing model M3 with M4 later on. This is not surprising, due to the exponentially increasing densities and temperatures towards the inner part of the envelope, which is expected to cause a strong absorption feature close to the systemic velocity. This absorption will conceal the contribution from a large scale surrounding cloud that is filtered out by the interferometer. To summarise, even if an extended component is added to the model, the central absorption will still be there, due to the presence of an infalling envelope.

It should be noted here, that the assumed velocity profile has a strong impact on the shape of the line profiles. The spatial resolution does not allow us to make a detailed analysis on the variation of the velocity of the gas with distance from the central source at the smallest scales (a few arc-seconds) and the data presented here does not show a clear trend that the velocity is increasing with increasing distance from the central source (Fig. 3). As previously mentioned, however, a pure Hubble-law velocity profile is favoured from the observations of H$_2$CO \citep{Oya:2014kx} and for that reason we assume that this is the case. 

Despite the fact that no obvious bow-shaped structures are present and that no emission at very high velocities is observed in this outflow, we also consider the case where spot shocks along the jet direction contribute to the observed emission line profiles. This case is included for completeness but is not likely, based on the very high densities and temperatures that are needed in the relatively confined regions, where shocks can be expected to be present. Although we cannot exclude the contribution from shocked regions completely, this component cannot make a large contribution to the observed emission in this particular source. In model M4, we consider a scenario where an outflow, an infalling envelope and shocked regions along the jet axis are included. The maximum velocity, at the apex of the bow-shocks is in the presented model set to, $\upsilon_{\rm{max}}$~=~20~\kmpers. Even though the densities and temperatures in the bow-shocks are quite high (1\texpo{8}~\cmthree\ and 1000~K, respectively), the contribution to the line profiles is still moderate, and it mostly has an effect at large distances from the central source. In M4, we have also excluded the cloud component to ease the comparison between the different models. 

To conclude, the M3 scenario can account for the main features in the line profiles both towards the central region and at offset positions in the outflow component. In addition, this model also reproduces the overall morphology of the CO emission.  
\section{Discussion}
\label{section:discussion}
\label{section:averyyoungoutflowsource}
\subsection{A very young outflow source traced by the $^{12}$CO emission?}
The outflow from \irasfemtontre\ was mapped in \cotretva\ using \apex\ \citep{van-Kempen:2009rt}. From the morphology of the red-shifted and blue-shifted outflow lobes, these authors concluded that the flow had a pole on geometry where the inclination angle with respect to the line of sight was small. The cause of this overlap was, however, likely the large beam size of the APEX telescope at these wavelengths. Inspection of the best-fit model (M3), presented in Sec.~\ref{section:radiativetransfermodeling}, convolved with the beam-size of \apex, reveal a similar \cotretva\ morphology, as was presented by these authors (their Fig.~13). To conclude, an outflow of short extent will appear to be pole-on, when observed with a large beam.

Recent observations with ALMA by \citet{Jorgensen:2013lr} and \citet{Oya:2014kx} revealed two clearly separated outflow lobes suggesting an inclination angle of 20\adeg\ with respect to the plane of the sky \citep{Oya:2014kx}, i.e., almost edge-on. From Figs.~\ref{fig:speciesmaps} \& \ref{fig:cochannelmap}, it is evident that the emission is confined to a region very close to the central source and that it has an edge on geometry. The absence of emission on large distances from \irasfemtontre\ ($>$15\asec), suggests that this outflow has a very young age. One obvious argument against this claim could be that the cloud is more dilute at larger distances, but inspection of the $^{13}$CO map presented in \citet{Tachihara:1996vn}, clearly shows that the cloud has a large extent in the east-west directions (\about30\amin). Also, there are no signs of Herbig-Haro objects on scales larger than 20\asec\ from the central source \citep{Heyer:1989lr}, supporting the interpretation that this outflow is very young. Assuming that the maximum radial velocity of the outflow is equal to the maximum observed velocity with respect to \vsys\ (i.e. \about8~\kmpers, which should be considered as a lower limit to the true maximum outflow velocity) and $i$~=~20\adeg, we estimate the dynamical time-scale of the flow to be of the order 500 years (see~Sec.~\ref{section:physicalproperties}). This puts \irasfemtontre\ in the same regime as other known outflow sources of very young age \citep[see e.g.][]{Richer:1990oa,Bourke:2005qy}.

\subsection{Episodic mass ejections}
As mentioned briefly already in Sec.~\ref{section:distributionofdetectedspecies}, recent observations with ALMA  \citep{Jorgensen:2013lr} show that HCO$^+$ is absent in the region closest to the protostar. These authors observed the isotopologue H$^{13}$CO$^+$ and suggest that this ion was destroyed by water vapor that was previously present on these scales. Such a scenario can also explain the emission of CH$_3$OH on small spatial scales and the extended observed carbon-chain chemistry. An increased water abundance is expected if the protostar underwent a recent burst in luminosity. As speculated by these authors, such a luminosity increase could be caused by a recent  increase in the infall rate towards the protostar, yielding increased shock chemistry in the outflow. From inspection of Fig.~\ref{fig:cochannelmap} it is evident that the emission, with a velocity offset of \about6~\kmpers\ with respect to \vsys, peak at four different locations in the outflow lobes (upper right panel). Two knots are visible on each side of the central source and the projected distance between each knot is \about5\asec. The more or less equal separation between the knots suggests that these features are due to periodic mass ejections, likely accompanied by periodic mass accretion events. The dynamical time-scale for these knots is \about100 years, consistent with the analysis presented in  \citet{Jorgensen:2013lr}, where the luminosity outburst are estimated to occur on a time-scale shorter than 100 -- 1000 years. The position velocity cut along the outflow direction (Fig.~\ref{fig:pv1}) clearly shows the episodic events and possibly also weak signs of acceleration, e.g., the "Hubble law" of outflows. A velocity cut along the presumed direction of the disk for $^{12}$CO and $^{13}$CO, does not reveal any signs of rotation. Perhaps the reason being the source is too young to have built up any appreciable disk mass on larger distances.

To summarise, the data suggest that the \irasfemtontre\ source is very young, likely younger than 1000 years. The emission at high velocities also indicates a shift in the outflow direction (see Fig.~\ref{fig:cochannelmap}). This could possibly be due to precession of the ejection axis and/or interaction between the outflow and the ambient gas. It is also not clear from this dataset whether \irasfemtontre\ is a single or binary source \citep[e.g.][]{Dunham:2014fk}, which can complicate the picture even further.    

\subsection{Physical properties of the outflowing gas}
\label{section:physicalproperties}
Since both $^{13}$CO and $^{12}$CO was mapped, and detected, in the outflowing gas, we can  estimate the opacity of the $^{12}$\cotvaett\ line. In the outflow positions where the $^{13}$CO emission peak (see Fig.~\ref{fig:speciesmaps}), and at the systemic velocity, the $^{12}$CO to $^{13}$CO ratio is close to 1, implying that the medium is optically thick. However, with increasing velocity, this ratio increases drastically. Unfortunately, the signal to noise ratio, for the high-velocity emission in the $^{13}$CO line is too low to obtain a firm value on the optical depth. Nevertheless, assuming that the line ratio is increasing with increasing velocity, we can give a lower limit to the $^{12}$CO to $^{13}$CO line ratio for the high velocity gas. For velocities 2~\kmpers\ offset from \vsys, this ratio is higher than 30 in the red-shifted outflow lobe, and higher than 20 in the blue shifted outflow lobe. As this is a lower limit in the velocity regime where $^{13}$CO is not detected, and since the line ratio increases dramatically with increasing velocity, we find it reasonable to assume that the lines are optically thin in the line wings. For the analysis presented here, we only take this optically thin regime into account. Thus, the mass estimates should be interpreted as lower limits to the true mass in the outflow lobes.

In the optically thin regime, the integrated emission in the line wings can be converted to a column density, assuming LTE conditions and using standard techniques \citep[see e.g.][]{Wilson:2009kx}. Since no other rotational lines of CO have been observed on these small scales, it is not possible to get a robust estimate on the excitation temperature along the outflow. If we assume a typical excitation temperature of 100~K in the outflow \citep[see e.g.][]{van-Kempen:2009rw}, and a CO/\htva\ ratio of \expo{-4} \citep[see e.g.][]{van-Dishoeck:1987mz}, the \htva\ column density is estimated at 3\texpo{20}~\cmtwo\ in the red-shifted and blue-shifted outflow lobes. Due to the uncertainty of this value, it is worth noting that the column density does not change by more than a factor of 5, when the excitation temperature is varied between 10 and 500 K. Given the extent ($L_{\rm{lobe}} \simeq\ 3700$ AU) and width (\about700~AU) of the flow, these numbers correspond to a total outflow mass of 3\texpo{-4}~\msun. The inferred value is consistent with the estimates presented by \citet[][$M_{lobe} \geq 4 \times 10^{-4} M_{\odot}$]{Dunham:2014fk} and \citet[][$M_{lobe} \simeq 3 \times 10^{-4} M_{\odot}$]{Yildiz:2015jk}, and it is small when compared to other Class~0 sources \citep[see e.g.][where estimates typically are higher than \expo{-4}~\msun]{Cabrit:1992lr,Wu:2004fk,Curtis:2010hc,Dunham:2014fk,Bjerkeli:2013fk}. The outflow mass is also very small compared to the protostellar mass, which has been estimated to 4\texpo{-2}~\msun\ \citep{Oya:2014kx}. This confirms that only a small amount of gas has been ejected out into the outflow and it supports the conclusion that the source is very young. 

The other characteristic flow parameters can be estimated from the flow extent, the mass and the velocity. First, the dynamical time scale of the flow is inferred from the extent and the maximum observed velocity in the red-shifted ($\upsilon_{\rm{max}} \simeq 6$~\kmpers) and blue-shifted ($\upsilon_{\rm{max}} \simeq 8$~\kmpers) outflow lobe, deprojected by the inclination angle with respect to the line of sight ($\bf{\it{t}_{dyn} = \it{L}_{lobe} \rm{cos(}\it{i}) / \upsilon_{\rm{max}}}$). The maximum deprojected red-shifted velocity is thus 18~\kmpers\ and the corresponding number for the blue-shifted flow is 23~\kmpers. This gives a dynamical time-scale of the outflow between 400 and 500 years. This is a factor of two lower than the value reported by \citet{Yildiz:2015jk}, based on APEX-CHAMP+ observations. When inferring the energetic parameters, we use the method suggested by \citet{Downes:2007lr}. For the momentum, the characteristic velocity, $\upsilon_{char}$ \citep{Lada:1996fk}, is used and no inclination correction is applied. To derive the characteristic velocity, we use the method described in \citet{Andre:1990fk}, where the intensity-weighted absolute velocity averaged over the mapped area is calculated. The derived flow momentum is 2\texpo{-4}~\msun~\kmpers\ and 4\texpo{-4}~\msun~\kmpers, in the red- and blue-shifted outflow lobe, respectively. For the kinetic energy, we use a factor of 5 when correcting for inclination \citep[][their Figs. 4 \& 5]{Downes:2007lr}, and we arrive at 2\texpo{40} erg and 5\texpo{40} erg, in the red- and blue-shifted lobe, respectively. The force of the outflow (momentum rate) is measured at \about1\texpo{-6}~\msun~\kmpers~yr$^{-1}$, i.e., slightly lower than the number presented in \citet{Yildiz:2015jk}. Compared to the results presented in that paper, also the mechanical luminosity is estimated at a slightly lower value, \about1\texpo{-3} \lsun, which can be compared to the bolometric luminosity of the source, 1.8~\lsun\ \citep{Jorgensen:2013lr}. Unfortunately, the limited spatial resolution of the dataset presented here, does not allow us to estimate the infall rate. A meaningful comparison with the mass-loss rate would, however, also be prohibited by the unknown size of the presumed protostellar disk. The mass-loss rate can readily be obtained from the momentum rate and an assumed velocity of the wind \citep[see e.g.][]{Goldsmith:1984fk}. Setting the wind velocity equal to the maximum flow velocity we derive a mass-loss rate, \about7\texpo{-8}~\msun~yr$^{-1}$. To conclude, the mass-loss rate, the mechanical luminosity, the momentum, the momentum rate, and the kinetic energy are all estimated at relatively low values. 

Compared to previous studies of outflows \citep[cf. e.g.][]{Wu:2004fk,Curtis:2010hc,Dunham:2014fk}, the values presented here fall at the lower end. In those papers, mass-loss rates are typically reported to be higher than \expo{-9}~\msun\ yr$^{-1}$ and the mechanical luminosity is for the bulk part of observed outflows higher than \expo{-3}~\lsun. Similarly are the momentum, momentum rates, and kinetic energies typically found to be higher than \expo{-3}~\msun~km~s$^{-1}$, \expo{-6}~\msun~km~s$^{-1}$~yr$^{-1}$, and \expo{41}~erg, respectively. It is therefore worth to note again that \irasfemtontre\ is a Class~0 source. Younger outflow sources are typically found to be more energetic than their evolved counterparts \citep[see e.g.][]{Bontemps:1996ys}. Why \irasfemtontre\ shows such characteristics as presented here is not entirely clear to us, but one reason may be that only a small amount of gas and dust up to now has been entrained by the outflow (hinted by the relatively low outflow mass). Another important aspect may be that \irasfemtontre\ is uncommon, in the sense that it has a highly variable accretion rate. If the protostar for long periods stay in a low accretion-rate mode, that will also have an impact on the energetics of the flow.
\begin{table}[t]  
\flushleft
\caption{Physical parameters of the outflow derived from the $^{12}$CO emission maps.}              
\label{table:limemodel}      
\resizebox{\hsize}{!}{
\begin{tabular}{l l l}          
  \hline\hline                        
  \noalign{\smallskip}
  \noalign{\smallskip}
  & \textit{Red lobe:} & \textit{Blue lobe:}\\
  \noalign{\smallskip}
 M$_{lobe}$ (\msun): 							& 2\texpo{-4} & 2\texpo{-4}\\
 L$_{lobe}$ (pc): 							&	8\texpo{-3} & 1\texpo{-2}\\
 R$_{lobe}$ (pc): 							& 	3\texpo{-3} & 4\texpo{-3}\\
 $\upsilon_{char}$  (km s$^{-1}$): 				&	5.1	& 7.4\\
 $\upsilon_{max}$  (km s$^{-1}$):				&	6.0 	& 8.0\\
 $t_{dyn}$ (yr):								&	4\texpo{2}	& 5\texpo{2}\\	
 Momentum (\msun\ km s$^{-1}$):				&	2\texpo{-4} & 4\texpo{-4}\\
 Kinetic energy (erg):							& 	2\texpo{40} & {5\texpo{40}}\\
 Momentum rate (\msun\ km s$^{-1}$ yr$^{-1}$):	&	4\texpo{-7} & 1\texpo{-6}\\
 Mechanical luminosity (\lsun):					&	3\texpo{-4} & 1\texpo{-3}\\
 Wind mass-loss rate (\msun\ yr$^{-1}$):			&	3\texpo{-8} & 4\texpo{-8}\\
  \noalign{\smallskip}
  \noalign{\smallskip}
  \hline                                             
  
\end{tabular}
}
\end{table}
\section{Conclusions}
We present maps of CO, HCO$^+$, N$_2$H$ ^+$ and N$_2$D$^+$ towards the \irasfemtontre\ region, observed with the Sub-Millimeter Array (SMA). These data were taken with a higher spatial resolution (2.6\asec\ -- 4.6\asec) than has previously been done for CO. From the analysis of these data and from comparison with recent ALMA observations, we conclude the following: 
\begin{itemize}
\item $^{12}$CO\,(2--1) and $^{13}$CO\,(2--1) are detected towards the \irasfemtontre\ region and are shown to have an outflow origin. C$^{18}$O\,(2--1) is detected towards the central region and only at blue-shifted velocities. HCO$^+$\,(3--2), and N$_2$H$^+$\,(3--2) is also tracing the outflow emission, while N$_2$D$^+$\,(3--2) is absent in the outflow and only detected to the Southwest. This suggests a non-outflow origin for this particular species. 
\item From analysis of the CO emission, where signs of episodic mass ejections are obvious, we conclude that the \irasfemtontre\ source is of a very young age, possibly younger than \about1000 years. 
\item The physical properties of the outflow, such as the mass, momentum, momentum rate, mechanical luminosity, kinetic energy and mass-loss rate are estimated at relatively low values. The data does not reveal any signs of a rotating disk like structure on these scales. 
\item A full 3D radiative transfer model of the system, using the Line Modeling Engine (LIME), explains the observations well. In order to reproduce the kinematical features in the CO line profiles as well as the morphology of the flow, three different components are needed, viz. a circumstellar envelope, a surrounding cloud and a wide-angle wind outflow. 
\end{itemize}
\label{section:conclusions}
\section*{}
      \small{\textit{Acknowledgements.} We thank the anonymous referee for a thorough report that greatly improved the quality of the paper. This research was supported by the Swedish research council (VR) through the contract 637-2013-472 to Per Bjerkeli and by the Lundbeck Foundation Junior Group Leader Fellowship to \mbox{Jes K. J\o rgensen}. Centre for Star and Planet Formation is funded by the Danish National Research Foundation. We also acknowledge support from National Science Foundation grant 1008800 and EU A-ERC grant 291141 CHEMPLAN.
     The Submillimeter Array is a joint project between the Smithsonian Astrophysical Observatory and the Academia Sinica Institute of Astronomy and Astrophysics and is funded by the Smithsonian Institution and the Academia Sinica.}
\bibliographystyle{aa}
\bibliography{papers}

\end{document}